\documentclass[review]{elsarticle}

\usepackage[utf8]{inputenc}
\usepackage[T1]{fontenc}
\usepackage{tgtermes}
\usepackage{lineno,hyperref,pifont,natbib,geometry,fleqn,graphicx,makeidx,fancyhdr,multirow,verbatim,amsmath,amsthm,amssymb,float,array,subfloat,epstopdf,subfig,upgreek,appendix,placeins}

\journal{Journal of Instrumentation}

\begin{document}

\begin{frontmatter}

\title{Modeling of surface-state induced inter-electrode isolation of $n$-on-$p$ devices in mixed-field and $\gamma$-irradiation environments}


\author[a]{N. Akchurin}
\author[a]{T. Peltola\corref{cor2}}
\ead{timo.peltola@ttu.edu}

\address[a]{Department of Physics and Astronomy, Texas Tech University, 1200 Memorial Circle, Lubbock, Texas, U.S.A.}
\cortext[cor2]{Corresponding author}




\begin{abstract}
\label{Abstract}
Position sensitive $n$-on-$p$ silicon sensors will be utilized in the tracker and in the High Granularity Calorimeter (HGCAL) of the Compact Muon Solenoid (CMS) experiment at High Luminosity Large Hadron Collider (HL-LHC).
%
The detrimental effect of the radiation-induced accumulation of
positive net oxide charge on position resolution in $n$-on-$p$ sensors has typically been countered by the application
of isolation implants like $p$-stop or $p$-spray between $n^+$-electrodes.
In addition to the positively charged layer inside the oxide and close to the Si/SiO$_2$-interface, surface damage introduced by ionizing radiation in SiO$_2$-passivated silicon particle detectors includes the accumulation of trapped-oxide-charge and interface traps.
%
%
%
A previous study of either n/$\gamma$ (mixed field)- or $\gamma$-irradiated Metal-Oxide-Semiconductor (MOS) capacitors 
showed evidence of substantially higher introduction rates of acceptor- and donor-type deep interface traps ($N_\textrm{it,acc/don}$) in mixed-field environment. In this work, an inter-pad and -strip 
resistance (or resistivity ($\rho_\textrm{int}$)) simulation study of $n$-on-$p$ sensors with and without $p$-stop isolation implants was conducted for both irradiation types.
Higher levels of $\rho_\textrm{int}$ showed correlation to higher densities of deep $N_\textrm{it,acc/don}$, with the inter-pad isolation performance of the mixed-field irradiated sensors becoming independent of the presence of $p$-stop implant between the $n^+$-electrodes 
up to about 100 kGy.  
%
The low introduction rates of deep $N_\textrm{it,acc/don}$ 
in $\gamma$-irradiated sensors 
resulted in high sensitivity of 
$\rho_\textrm{int}$ to the presence and peak doping of $p$-stop 
above the lowest dose of about 7 kGy in the study. 
%
%
As a consequence of the advantageous influence of radiation-induced accumulation of deep $N_\textrm{it}$ on the inter-electrode isolation, position sensitive $n$-on-$p$ sensors without isolation implants may be considered for future HEP-experiments 
where the radiation is largely due to hadrons.
%

\end{abstract}

%
\end{frontmatter}


\newpage
\section{Introduction}
\label{Introduction}
In the position sensitive $n$-on-$p$ particle detectors that will be utilized in the experiments of the High-Luminosity Large Hadron Collider (HL-LHC) like ATLAS and 
CMS, a parameter of particular concern is the preservation of position resolution in extreme radiation environments. 
This is due to the disadvantage of the $n$-on-$p$ sensors where the segmented $n^+$-electrodes can become shorted even when operated at high reverse bias voltages. The issue is caused by the radiation-induced accumulation of positive charge inside the passivation SiO$_2$ and close to the Si/SiO$_2$-interface 
that can attract electrons to the 
interface in the inter-electrode gap 
to the extent that creates a conduction channel 
and compromises the inter-electrode
isolation 
of the sensor. 
This is typically 
alleviated by placing an additional highly $p$-doped implant between the $n^+$-electrodes that breaks the conduction channel, due to its negative space charge, by suppressing the electron density
at the vicinity of Si/SiO$_2$-interface. The two solutions commonly utilized to produce the isolation implants involve localized ($p$-stop) and uniformly distributed ($p$-spray) doping concentrations along the surface in the inter-electrode gap \cite{Kemmer1993,Richter1996,Iwata1998,Verzellesi2000,Piemonte2006,Unno2013,Printz2016}. 
The $p$-spray approach adds one and $p$-stop two more lithography steps and additional ion implantations \cite{Harkonen2007b,Pellegrini2007} 
to the processing steps of a conventional $p$-on-$n$ sensor, increasing the production cost of $n$-on-$p$ sensors.
%
Additionally, since high doping concentrations in the isolation implants 
have been observed to increase the probability of discharges or avalanche effects due to excessive localized electric fields \cite{Adam2017,Printz2016,Dalal2014b}, careful tuning of the processing parameters is required.

It should be noted that due to the positive oxide charge of SiO$_2$ at the Si/SiO$_2$-interface ($N_\textrm{ox}$), the $n^+$-electrodes without isolation implants are shorted by default already before irradiation, but as shown in \cite{Peltola2024}, become isolated by the application of reverse bias voltage above the \textit{threshold voltage of inter-electrode isolation} ($V_\textrm{th,iso}$) (e.g., $V_\textrm{th,iso}\leq100$ V for $N_\textrm{ox}\leq7\times10^{10}~\textrm{cm}^{-2}$).

Surface damage in SiO$_2$-passivated devices induced by ionizing radiation (charged particles, X-rays or $\gamma$s) 
consists of the accumulation of a positively charged layer (fixed oxide charge density $N_\textrm{f}$ that does not move or exchange charge with Si), trapped-oxide-charge and interface traps (or surface states $N_\textrm{it}$), along with mobile-ionic-charge ($N_\textrm{M}$) inside the oxide and close to the interface with silicon bulk. The mechanisms 
of surface-damage accumulation 
are described extensively in \cite{Nicollian1982,Dressen1989,Oldham1999,Schwank2008,Zhang2013}. 

A previous study on inter-electrode resistance ($R_\textrm{int}$) performances of $n$-on-$p$ sensors with $p$-stop or $p$-spray isolation implants irradiated either by protons or neutrons and $\gamma$s 
or the combination of the two up to estimated Total Ionizing Doses (TID) of about 1.5--2 MGy\footnote{Estimated by 1 kGy and 145 kGy per $1\times10^{14}~\textrm{n}_\textrm{eq}\textrm{cm}^{-2}$ for neutrons \cite{Mandic2004} and protons \cite{Gosewich2021_J}, respectively, for the irradiation facilities in \cite{Adam2017}.} showed high inter-strip isolation levels that, after the initial drop from the pre-irradiated level, remained essentially constant throughout the studied TID-range \cite{Adam2017}. However, corresponding studies on X-ray irradiated $n$-on-$p$ sensors with $p$-stop or $p$-spray \cite{Moscatelli:2017sps} or solely $p$-stop \cite{Mariani2020} isolation implants displayed significant dose dependences, where $R_\textrm{int}$-values decreased by about two orders of magnitude between 0.5--100 kGy in \cite{Moscatelli:2017sps} and 50--700 kGy in \cite{Mariani2020}. Additionally, when X-ray irradiations were complemented by n/$\gamma$-irradiations \cite{Mariani2020}, the dose (or TID) dependence disappeared and high isolation levels were maintained for all TIDs in the study. Significantly, the higher-level inter-electrode isolation 
for irradiation types involving hadrons were 
found also for $n$-on-$p$ sensors without isolation implants in \cite{Gosewich2021_J}. In this study, the sensors irradiated by protons or n/$\gamma$s to estimated TIDs of 870 and 6 kGy, respectively, displayed substantially higher $R_\textrm{int}$-levels to a sensor X-ray irradiated to a dose of 3 kGy. These observations suggest that hadron or mixed-field irradiation induces higher 
rates of surface states with beneficial influence on inter-electrode isolation compared to X-ray irradiation.

This investigation focuses on the correlation between the introduction rates of deep acceptor- and donor-type interface traps ($N_\textrm{it,acc/don}$) and inter-electrode isolation in mixed-field or $\gamma$-irradiated $n$-on-$p$ sensors with and without $p$-stop isolation implants.
The study presented here is a 
continuation of the investigation reported in \cite{Peltola2023}, where the parameters of the components of the radiation accumulated net 
$N_\textrm{ox}$ $-N_\textrm{f}$ and deep $N_\textrm{it,acc/don}$ (no evidence of $N_\textrm{M}$ was detected)$-$ were extracted by reproducing by simulation the $CV$-characteristics of n/$\gamma$ (i.e., a mixed field)- or $\gamma$-irradiated MOS-capacitors. 
Motivated by the observed significant differences in the introduction rates of $N_\textrm{it,acc/don}$, 
the investigation was extended to a comparison of the simulated $R_\textrm{int}$ (or resistivity ($\rho_\textrm{int}$)) 
performances between the two irradiation types. 
By applying the extracted $N_\textrm{f}$ and $N_\textrm{it,acc/don}$ at the Si/SiO$_2$-interface between $n^+$-electrodes of $n$-on-$p$ sensor structures, 
$\rho_\textrm{int}$-simulations were initiated for the mixed field in the previous study and completed for both irradiation types in this investigation. The results 
were produced by sensor structures with dimensions and parameters of CMS High Granularity Calorimeter (HGCAL) \cite{Phase2} pad sensors and test-strips.

The paper is arranged 
by first 
discussing in Section~\ref{TdS} the observations on oxide-charge and surface-state accumulation with dose ($D$), based on the results extracted in ref. \cite{Peltola2023}. 
Next, the 
simulation setup, modeled sensor structures and methods used to derive inter-electrode resistivity 
are described in Section~\ref{Msetup}. 
Inter-electrode isolation results in Section~\ref{Results} start with simulated $\rho_\textrm{int}$ performance of a pre-irradiated sensor with and without $p$-stop in Section~\ref{preIrr}. Corresponding results for mixed-field and $\gamma$-irradiated sensors are presented in Section~\ref{MF_gammas}, followed by the comparison between measured and simulated $\rho_\textrm{int}$ performances of X-ray and $\gamma$-irradiated sensors 
in Section~\ref{StripSensors}.
%
Finally, the results are discussed in Section~\ref{Discussion}, while summary and conclusions 
are given in Section~\ref{Summary}. 
\section{Observations on oxide-charge and surface-state accumulation with dose}
\label{TdS}
%
%
The experimental and simulation study presented in ref. \cite{Peltola2023} of MOS-capacitors n/$\gamma$-irradiated 
at the Rhode Island Nuclear Science Center\footnote{http://www.rinsc.ri.gov/} (RINSC) and UC Davis McClellan Nuclear Research Center\footnote{https://mnrc.ucdavis.edu/} (MNRC) reactors or $\gamma$-irradiated at Sandia National Laboratories Gamma Irradiation Facility\footnote{https://www.sandia.gov/research/gamma-irradiation-facility-and-low-dose-rate-irradiation-facility/} (GIF) led to the extracted surface-state parameters in Table~\ref{tabNit} with densities presented in Table~\ref{table_Simulated}. This process was carried out by tuning the parameters in a Technology Computer-Aided Design (TCAD)-simulation until close agreement with measured $CV$-characteristics of the irradiated MOS-capacitors was reached.

The 
TIDs in Table~\ref{table_Simulated} for MNRC were determined by Monte Carlo N-Particle Transport (MCNP) simulations, which are in line with the reported TID-rate of $1~\textrm{kGy}$ per $1\times10^{14}~\textrm{n}_\textrm{eq}/\textrm{cm}^{2}$ at Jo{\v{z}}ef Stefan Institute reactor facility (JSI) \cite{Ravnik2003,Mandic2004} $-$a site widely utilized in the Si-device R\&D within 
CMS and ATLAS$-$ that has a similar TRIGA Mark II-design to the MNRC reactor. Since reactor-specific TID-estimates were not available for RINSC and due to the similar open-pool-reactor types of all three facilities with comparable power between RINSC and MNRC (2.0 and 2.3 MW, respectively), the aforementioned TID-rate was considered as a reasonable estimate also for RINSC TIDs, shown in Table~\ref{table_Simulated}.
  
%
\begin{table*}[!t]
\centering
\caption{The simulation input parameters of radiation-induced 
$N_\textrm{it}$ from ref. \cite{Peltola2023}. 
$E_\textrm{a,V,C}$ are the activation energy, 
valence band and conduction band energies, respectively, while $\sigma_\textrm{e,h}$ are the electron and hole trapping cross sections, respectively.} 
\label{tabNit}
\begin{tabular}{|c|c|c|c|}
    \hline
    {\bf $N_\textrm{it}$ type} & {\bf $E_\textrm{a}$} \textrm{[eV]} & {\bf $\sigma_\textrm{e,h}$} \textrm{[cm$^{2}$]} & 
{\bf Density} \textrm{[cm$^{-2}$]}\\
    \hline
    Deep donor ($N_\textrm{it,don}$) & $E_\textrm{V}+0.65$ & $1\times10^{-15}$ & see column 5 in Table~\ref{table_Simulated}\\
    Deep acceptor ($N_\textrm{it,acc}$) & $E_\textrm{C}-0.60$ & $1\times10^{-15}$ & see column 6 in Table~\ref{table_Simulated}\\
    \hline
\end{tabular}
\end{table*}
%
\begin{table*}[!t]
\centering
\caption{
Neutron fluences, TIDs, $CV$-sweeps initialized from 
inversion (Inv., minority carriers at the Si/SiO$_2$-interface) and accumulation (Acc., majority carriers at the interface) modes of MOS-capacitor operation. 
Also given are the TCAD-simulation input densities for fixed oxide charge ($N_\textrm{f}$) and donor- and acceptor-type interface traps ($N_\textrm{it,don}$ \& $N_\textrm{it,acc}$, respectively) from Table~\ref{tabNit} to reproduce measured $CV$-characteristics of irradiated MOS-capacitors in ref. \cite{Peltola2023}. 
$^\ast$) MCNP-simulated TIDs, $^\top$) Estimated $\textrm{TID}\approx1~\textrm{kGy}$ per $1\times10^{14}~\textrm{n}_\textrm{eq}/\textrm{cm}^{2}$ \cite{Ravnik2003,Mandic2004}. $\textrm{Fluence}=0$ indicates the $\gamma$-irradiated samples.
}
\label{table_Simulated}
\begin{tabular}{|c|c|c|c|c|c|}
\hline 
\multirow{2}{*}{{\bf Fluence}} & \multirow{2}{*}{{\bf TID/Dose}} & \multirow{2}{*}{{\bf $CV$-sweep}} & \multirow{2}{*}{{\bf TCAD $N_\textrm{f}$}} & \multirow{2}{*}{{\bf TCAD $N_\textrm{it,don}$}} & \multirow{2}{*}{{\bf TCAD $N_\textrm{it,acc}$}}\\[0.9mm]%
[$\times10^{15}~\textrm{n}_\textrm{eq}/\textrm{cm}^{2}$] & [kGy] & { } & [$\times10^{12}~\textrm{cm}^{-2}$] & [$\times10^{12}~\textrm{cm}^{-2}$] & [$\times10^{12}~\textrm{cm}^{-2}$]\\
\hline
$0.35\pm0.04$ & 
$3.5\pm0.3^\top$ & Inv./Acc. & 0.68 & 1.11 & 1.08\\
\hline 
$0.61\pm0.05$ & 
$7.1\pm0.6^\ast$ & Inv./Acc. & 0.77 & 1.65 & 1.59\\
\hline
$2.35\pm0.19$ & 
$23.5\pm1.9^\top$ & Inv./Acc. & 1.00 & 2.30 & 2.60\\
\hline
$6.6\pm0.7$ & 
$64\pm7^\top$ & Inv./Acc. & 1.75 & 3.25 & 4.70\\
\hline
$9.3\pm1.1$ & 
$90\pm11^\ast$ & Inv./Acc. & 2.20 & 4.15 & 4.35\\
\hline
\multirow{2}{*}{0} & \multirow{2}{*}{$7.0\pm0.4$} & 
Inv. & 0.79 & 1.92 & 1.10\\ 
& & Acc. & 0.79 & 1.65 & 1.10\\
\hline
\multirow{2}{*}{0} & \multirow{2}{*}{$23.0\pm1.2$} & 
Inv. & 1.20 & 1.70 & 1.15\\
& & Acc. & 1.20 & 2.11 & 1.00\\
\hline
\multirow{2}{*}{0} & \multirow{2}{*}{$90\pm5$} & 
Inv. & 2.20 & 1.05 & 1.50\\
& & Acc. & 2.00 & 0.50 & 1.00\\
\hline
\end{tabular}
\end{table*}
\begin{linenomath}
After irradiation, the net oxide charge density at the Si/SiO$_2$-interface can be described as \cite{Peltola2023}
\begin{equation}\label{eq1}
Q_\textrm{ox}={e}N_\textrm{ox}=e(N_\textrm{f}+{a}N_\textrm{it,don}-{b}N_\textrm{it,acc}),
\end{equation}
where $e$ is the elementary charge and $a$ and $b$ are the fractions of 
occupied $N_\textrm{it,don}$ and $N_\textrm{it,acc}$, respectively. As shown in ref. \cite{Peltola2023}, since the values of $a$ and $b$ change with applied gate voltage ($V_\textrm{gate}$) 
$Q_\textrm{ox}$ in Eq.~\ref{eq1} 
can experience a polarity reversal with $V_\textrm{gate}$. 
\end{linenomath}

The differences in tuned values of $N_\textrm{f}$ and $N_\textrm{it,don/acc}$ for $\gamma$-irradiated MOS-capacitors between $CV$-sweeps initialized either from accumulation- or inversion-regions are evident in Table~\ref{table_Simulated}. This is due to observed hysteresis in the respective $CV$-curves, which was found to be caused by the differences in the fractions of occupied $N_\textrm{it,don/acc}$ depending on the polarity of the initializing $V_\textrm{gate}$ \cite{Peltola2023} (e.g., for $CV$-sweep starting from negative $V_\textrm{gate}$ a larger fraction of $N_\textrm{it,don}$ will be occupied, resulting in a higher value of $N_\textrm{ox}$ in Eq.~\ref{eq1}). No hysteresis was observed in the $CV$-characteristics of mixed-field irradiated MOS-capacitors, resulting in a single set of $N_\textrm{f}$ and $N_\textrm{it,don/acc}$ for a given TID in Table~\ref{table_Simulated}.
Summary of MOS-capacitor modes of operation is given in~\ref{App1}.

When the densities of the three components of $N_\textrm{ox}$ from Table~\ref{table_Simulated} are plotted as a function of dose in Figure~\ref{surfStatDens}, it is evident that while $N_\textrm{f}$-evolution is essentially equal for both of the irradiation types, $N_\textrm{it}$ introduction rate for both donors and acceptors is substantially higher for the mixed-field environment. This difference in introduction rates of $N_\textrm{it}$ could explain the previously reported 
better inter-strip isolation performance for mixed-field and hadron-irradiated $n$-on-$p$ strip-sensors with or without isolation implants compared to the X-ray irradiated ones \cite{Gosewich2021_J}, due to the beneficial impact of
radiation-induced accumulation of deep interface traps on the inter-electrode resistance ($R_\textrm{int}$) \cite{Peltola2023}.
As discussed in ref. \cite{Peltola2023}, the 
frequency- and temperature ($f,T$) dependence in the $CV$-characterizations introduced by the accumulation of $N_\textrm{it}$ \cite{Nicollian1982} 
could 
add a level of arbitrariness to the values of simulation extracted $N_\textrm{f}$ and $N_\textrm{it,acc/don}$. 
Comparison with $N_\textrm{f}$ and $N_\textrm{it,acc}$ 
extracted from capacitance/conductance-voltage 
and thermal dielectric relaxation current 
measurements of X-ray irradiated MOS-capacitors $-$considering the results for 700- and 750-nm-thick SiO$_2$ and $\langle100\rangle$ crystal orientation Si-bulk in Figures 9.6 and 9.7 in ref. \cite{Zhang2013}$-$ shows similar 
densities and 
dose dependencies with the $\gamma$-irradiation results in Table~\ref{table_Simulated} and Figure~\ref{surfStatDens}. 
Further comparison with 
measured $N_\textrm{ox}$ and $N_\textrm{it}$ results 
reported in \cite{Moscatelli:2017sps} for MOS-capacitors (with 650-nm-thick SiO$_2$ and $\langle100\rangle$ Si-bulk) X-ray irradiated up to 100 kGy shows again similar values with $\gamma$-irradiation accumulated $N_\textrm{f}$ and $N_\textrm{it,acc}$ presented here. 
%
\begin{figure*}
     \centering
    \includegraphics[width=.7\textwidth]{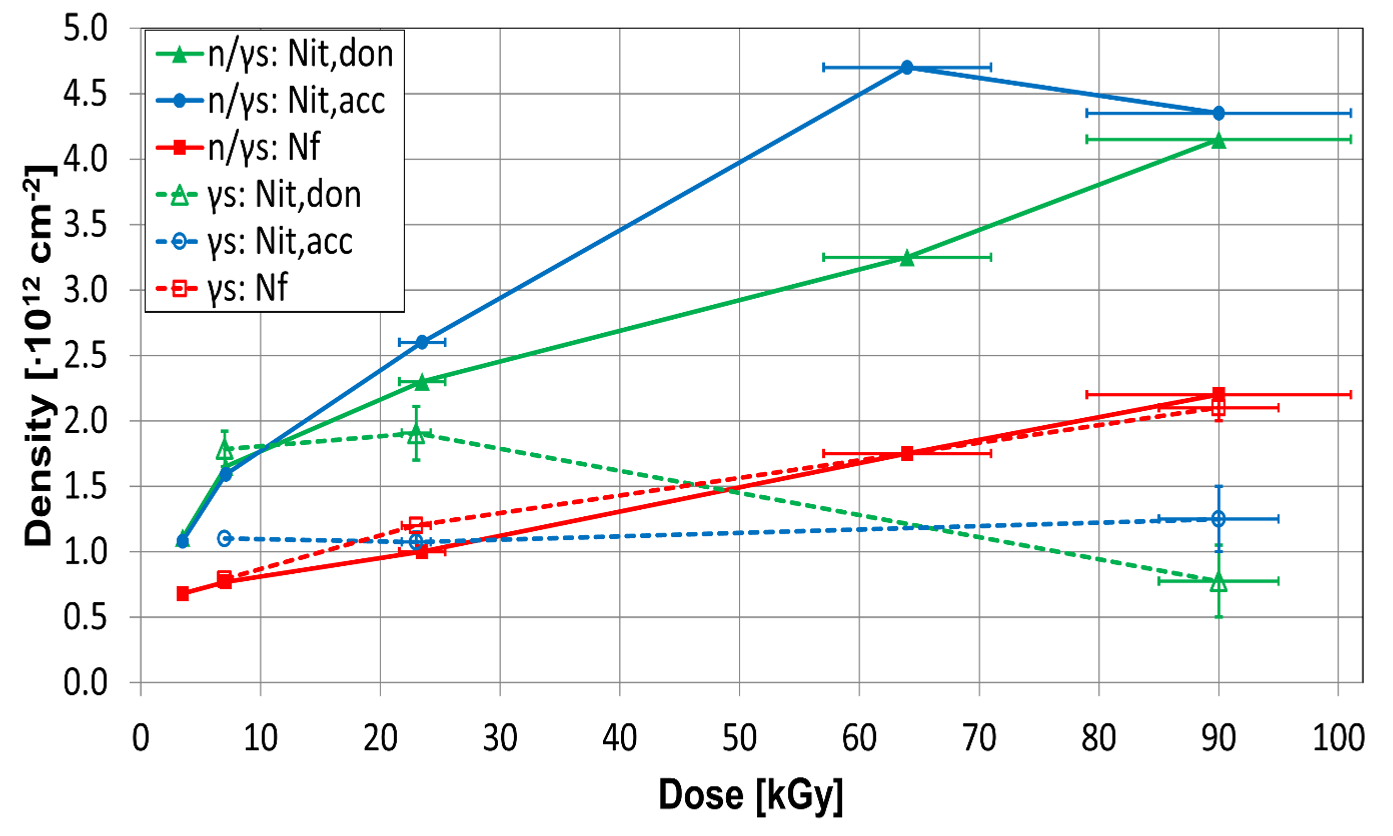}\label{surfStatDens}
    \caption{\small Simulated evolution of $N_\textrm{f}$ and surface-state densities ($N_\textrm{it,acc/don}$) with dose 
at room temperature and frequencies between $1-9~\textrm{kHz}$ from Table~\ref{table_Simulated} for mixed-field (`n/$\gamma$s') and $\gamma$-irradiations. For 
clarity, mean values between densities extracted from $CV$-sweeps initialized either from accumulation- or inversion-regions of the $\gamma$-irradiated MOS-capacitors are considered. 
%
}
\label{surfStatDens}
\end{figure*}
\section{Simulation setup and methods}
\label{Msetup}
\begin{figure}[htb!]
\centering
\subfloat[]{\includegraphics[width=.6\textwidth]{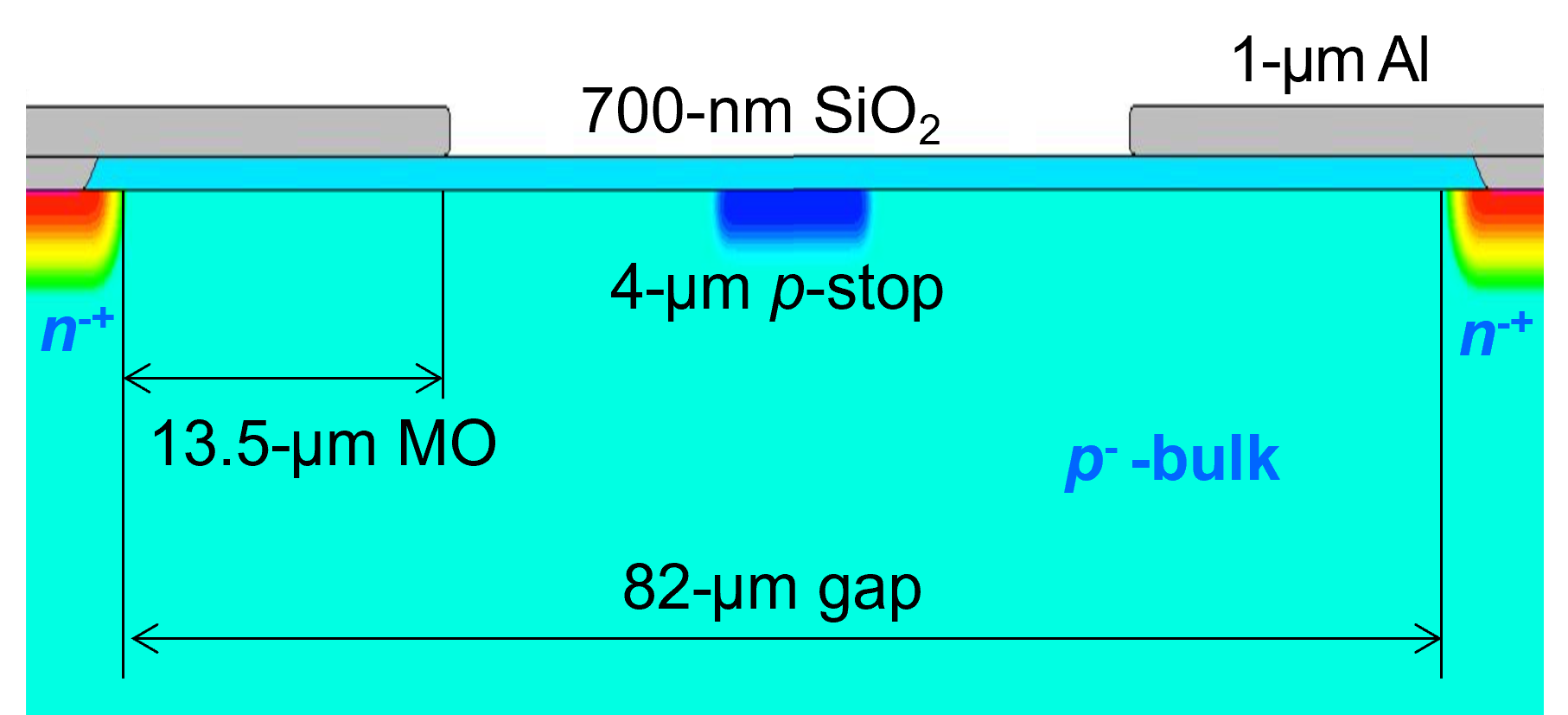}\label{RintDevice1}}\hspace{2mm}%
\subfloat[]{\includegraphics[width=.34\textwidth]{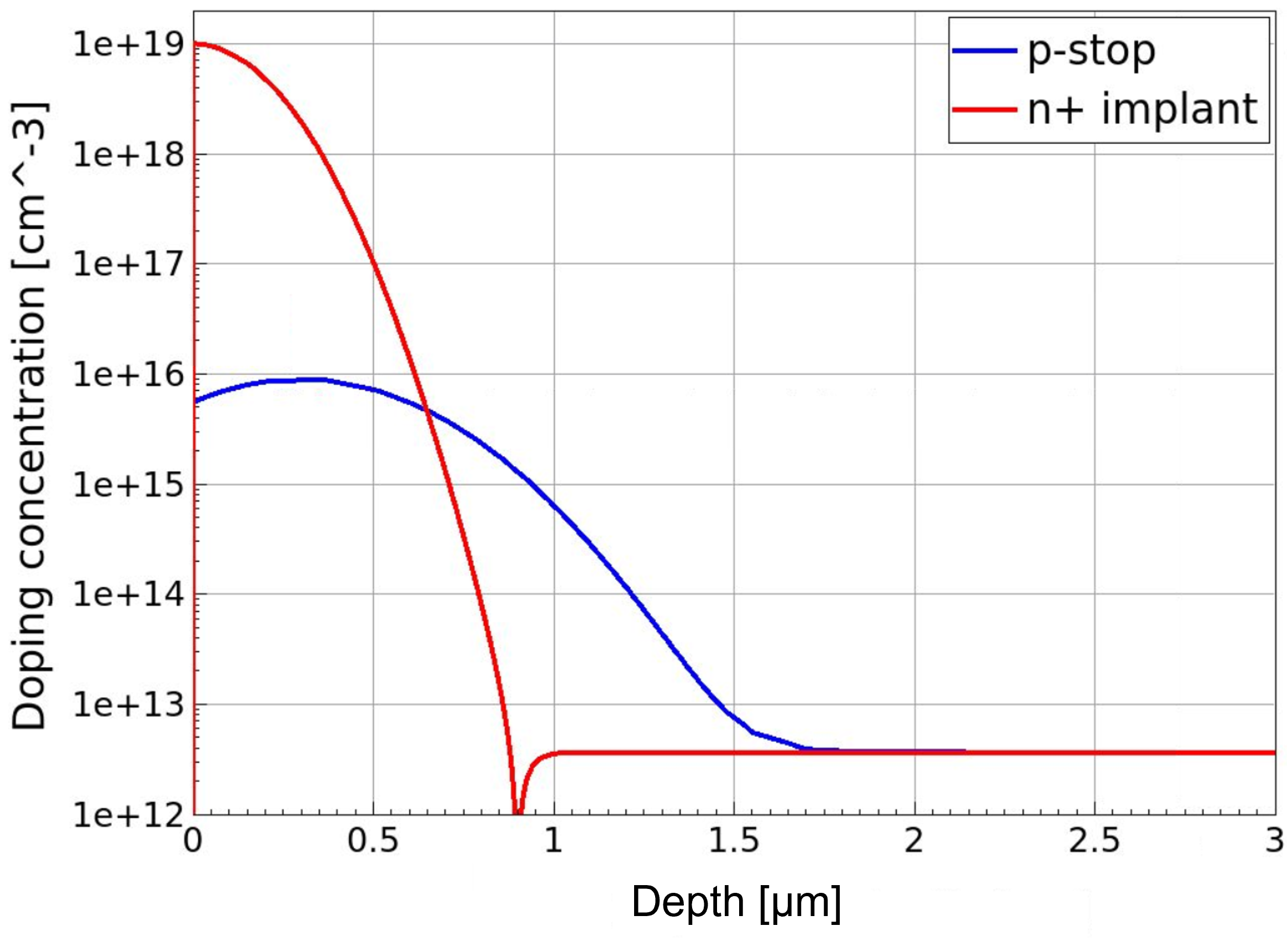}\label{dopeProfiles}}\\
\subfloat[]{\includegraphics[width=.6\textwidth]{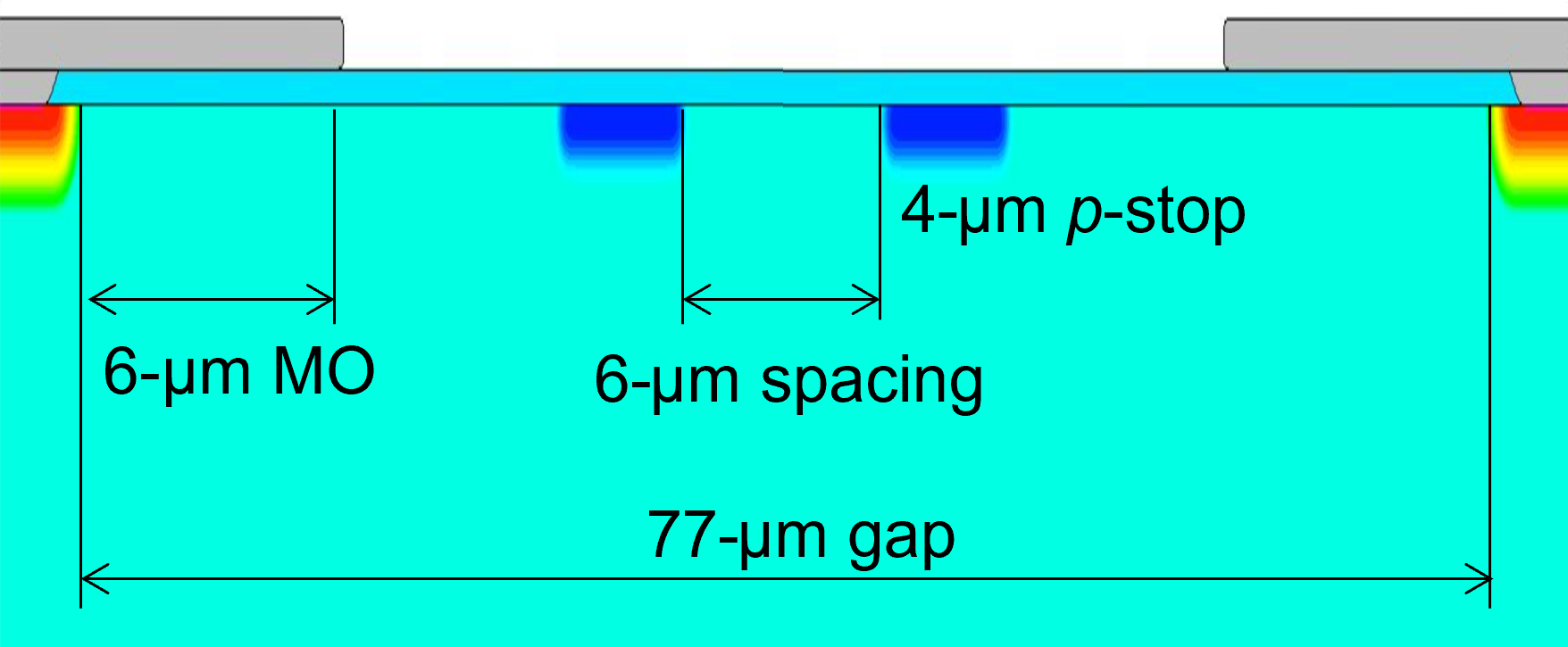}\label{RintDevice2}}
\caption{\small 2D-device structures implemented for the $R_\textrm{int}$-simulations. (a) Inter-pad region to 10-$\upmu\textrm{m}$ depth from the front surface of a DC-coupled 
300-$\upmu\textrm{m}$-thick $n$-on-$p$ pad-sensor with a `common' $p$-stop. MO = Metal-Overhang. A 700-nm-deep $p^+$-blocking contact with 1-$\upmu\textrm{m}$-thick uniform Al-layer was applied on the backplane of the sensor. (b) Doping profiles of the $n^+$-implants and $p$-stop with $N_\textrm{ps}=9.0\times10^{15}~\textrm{cm}^{-3}$. (c) Inter-strip region of a DC-coupled 300-$\upmu\textrm{m}$-thick $n$-on-$p$ strip-sensor with `individual' $p$-stops. Features different from Figure~\ref{RintDevice1} are indicated. Doping profiles are identical to Figure~\ref{dopeProfiles}, except for $N_\textrm{ps}=1.5\times10^{16}~\textrm{cm}^{-3}$.
%
}
\label{HGCALdevice}
\end{figure}

The 2D device-simulations 
were carried out using the Synopsys Sentaurus\footnote{http://www.synopsys.com} finite-element 
TCAD software framework. The simulations apply Neumann boundary and oxide-semiconductor jump conditions (potential jump across the material interface due to dipole layers of immobile charges), as well as Dirichlet boundary conditions for
carrier densities and AC potential at Ohmic contacts that are used to excite the system. 
For the 
$R_\textrm{int}$-simulations, device-structures displayed in Figure~\ref{HGCALdevice} were implemented. The device in Figure~\ref{RintDevice1} features the surface design dimensions between the cells of an HGCAL multi-channel sensor, while the device in Figure~\ref{RintDevice2} reproduces the surface design of HGCAL test-strips. For the sensor-structure in Figure~\ref{RintDevice1}, designs both with and without $p$-stop isolation implant were included in the simulations.
%
Results from Spreading Resistance Profiling (SRP) measurements of the sensors carried out within HGCAL-community \cite{Hinger2021} were used as an input for the doping profiles in Figure~\ref{dopeProfiles}, while the passivation oxide thickness ($t_\textrm{ox}$) input was provided by the data from $CV$-characterizations of MOS-capacitors from HGCAL-sensor wafers.
The bulk properties 
leakage current ($I_\textrm{leak}$) and full depletion voltage ($V_\textrm{fb}$) were set to match the measurements by tuning the charge carrier lifetimes ($\tau_\textrm{e.h}$) and bulk doping ($N_\textrm{B}$), respectively, while the crystal orientation $\langle100\rangle$ of the HGCAL sensor substrates was 
repeated in the bulk of the modeled devices. 
The components of radiation-induced surface damage, $N_\textrm{f}$ and $N_\textrm{it}$ from Table~\ref{table_Simulated}, were implemented 
at the Si/SiO$_2$-interface with a uniform distribution along the interface, which is an approximation of the real 
distribution that includes nonuniformities \cite{Nicollian1982}. 

%
%
%
The simulated $R_\textrm{int}$ was realized by applying a reverse bias-voltage ($V_\textrm{bias}$) from the backplane contact of the sensor, while either a low voltage $V_\textrm{int}$ or no voltage is applied from one of the $n^+$-electrodes and the resulting currents are observed on the adjacent electrode. $R_\textrm{int}$ is then determined as
\begin{linenomath}
\begin{equation}\label{eq2}
R_\textrm{int}=\frac{V_\textrm{int}-0~\textrm{V}}{I_\textrm{leak}(V_\textrm{int})-I_\textrm{leak}(0~\textrm{V})},
\end{equation}
%
by extracting the bulk-generated current $I_\textrm{leak}(0~\textrm{V})$ from the surface current.
\end{linenomath}
The method yields identical results to the approach where at a discrete value of $V_\textrm{bias}$, $I_\textrm{leak}$ is measured at multiple $V_\textrm{int}$ (typically voltages between $\pm5$ V to ensure linearity of $I_\textrm{leak}$($V_\textrm{int}$)) and $R_\textrm{int}$ is extracted from the slope of $I_\textrm{leak}$($V_\textrm{int}$)-plot. 
However, this procedure was chosen 
due to being substantially faster and less demanding on computing capacity since 
$R_\textrm{int}$ for the full $V_\textrm{bias}$-range is given 
directly by the simulation. 
For straightforward comparison with sensors of varied geometries, $R_\textrm{int}$ is normalized to 
$\rho_\textrm{int}$ by the cross-sectional area $A$ and the length $l$ (82 and 77 $\upmu\textrm{m}$ in Figures~\ref{RintDevice1} and~\ref{RintDevice2}, respectively) of the resistor as \cite{Knoll2000}
\begin{linenomath}
\begin{equation}\label{eq3}
\rho_\textrm{int}=R_\textrm{int}\frac{A}{l}=R_\textrm{int}\frac{w{\cdot}d}{l},
\end{equation}
%
where $w$ is the width of the electrode (i.e., length of the strip for strip-sensors) and $d$ is the depth of the $n^+$-implant (2 $\upmu\textrm{m}$ in Figure~\ref{dopeProfiles}).
\end{linenomath}

%

%
\section{Inter-electrode isolation results}
\label{Results}
The simulations in Sections~\ref{preIrr}--\ref{StripSensors} were carried out with the sensor-structure in Figure~\ref{RintDevice1}, while 
Section~\ref{StripSensors} also applied the strip-sensor structure in Figure~\ref{RintDevice2}.
\subsection{Simulated: Pre-irradiated sensor}
\label{preIrr}
\begin{figure*}
     \centering
    \includegraphics[width=.7\textwidth]{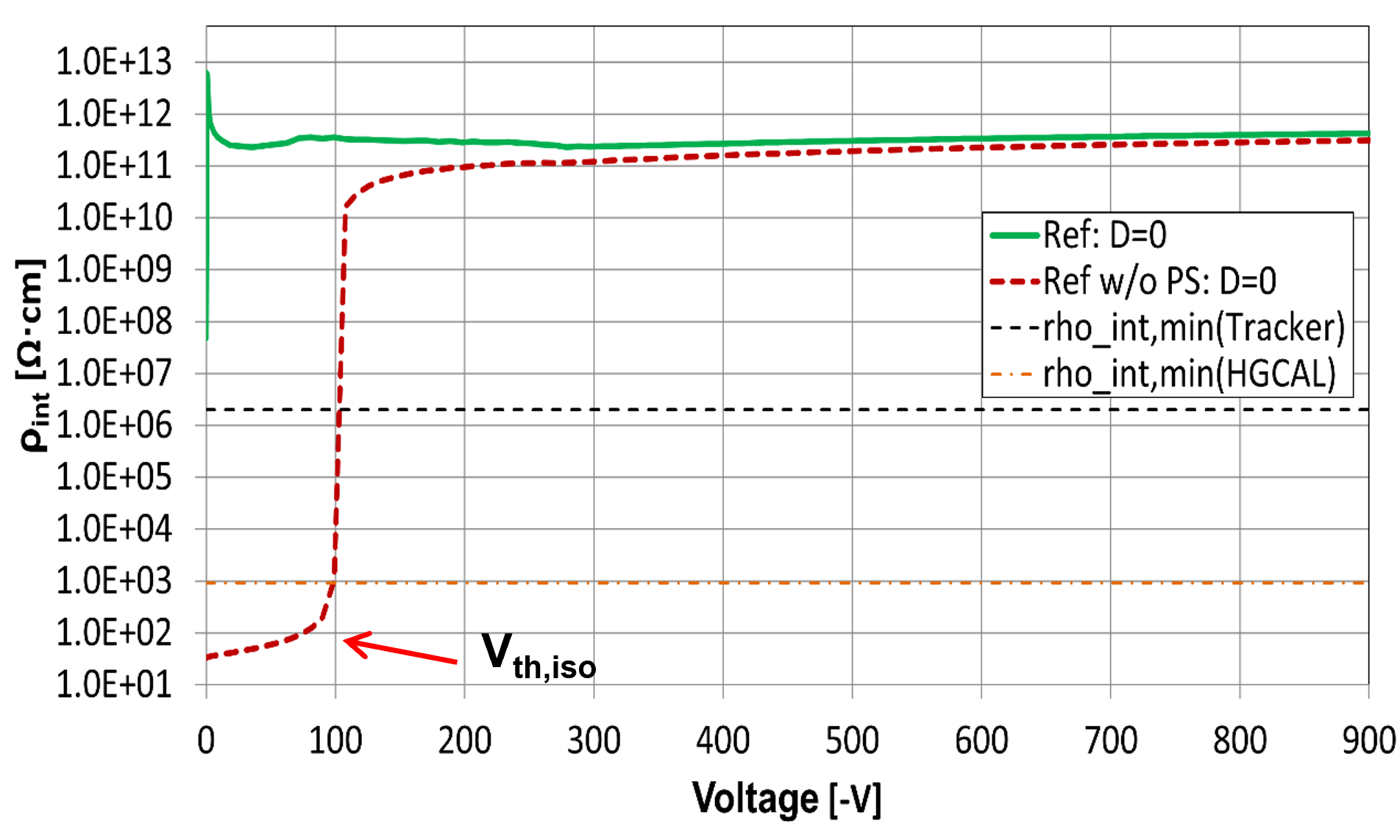}
    \caption{\small Simulated inter-electrode resistivity ($\rho_\textrm{int}$) for a pre-irradiated reference sensor at $T=293~\textrm{K}$. Si/SiO$_2$-interface applied $N_\textrm{ox}~{\cong}~N_\textrm{f}=6.8\times10^{10}~\textrm{cm}^{-2}$. Threshold voltage 
of inter-electrode isolation ($V_\textrm{th,iso}$) for the sensor without a $p$-stop implant (`w/o PS') is indicated, as well as estimates for the lower limits of sufficient inter-electrode isolation for the Tracker ($2.0~\textrm{M}\Omega\cdot\textrm{cm}$) and HGCAL ($0.9~\textrm{k}\Omega\cdot\textrm{cm}$) sensors. 
}
\label{Rint_REF}
\end{figure*}

Figure~\ref{Rint_REF} displays the $\rho_\textrm{int}$-results for a pre-irradiated reference sensor with and without $p$-stop isolation implant. As shown in ref. \cite{Peltola2023}, before irradiation $N_\textrm{ox}~{\cong}~N_\textrm{f}$ which was set to $6.8\times10^{10}~\textrm{cm}^{-2}$ according to field-region $N_\textrm{ox}$ results extracted from HGCAL sensors with oxide quality type `C' in ref. \cite{Peltola2024}. 

The dashed black and orange lines in Figure~\ref{Rint_REF} indicate the conservative estimates for the lower limits of sufficient inter-strip resistivity ($\rho_\textrm{int,min}$) for the 
CMS Tracker and HGCAL sensors, respectively. 
The difference between the two limits is due to the different limiting resistances in the biasing ($R_\textrm{bias}\approx2~\textrm{M}\Omega$ \cite{Hartmann2017}) and readout (preamplifier input impedance of a few hundred $\Omega$s) circuits of the AC-coupled Tracker and DC-coupled HGCAL sensors, respectively. The values of $\rho_\textrm{int,min}$ are determined by requiring $R_\textrm{int,min}$ to be about two orders of magnitude above the limiting circuit resistances ($\rho_\textrm{int,min}$ for the Tracker strips with $l=75~\upmu\textrm{m}$ and $w=1.55~\textrm{cm}$ is the corresponding value to the limit $R_\textrm{int,min}\times\textrm{strip length}=100~\textrm{M}\Omega\cdot\textrm{cm}$ used in refs. \cite{Gosewich2021_J,Gosewich2021}, while $\rho_\textrm{int,min}$ for the HGCAL is a current estimation 
applied for this study). Even though all simulations in this study were carried out for DC-coupled HGCAL sensors in Figure~\ref{HGCALdevice} (simulated $\rho_\textrm{int}$ is identical for AC- and DC-coupled devices) both $\rho_\textrm{int,min}$ limits are included in the following results as useful benchmarks. 
%

The threshold voltage of inter-electrode isolation ($V_\textrm{th,iso}$) \cite{Peltola2024} in Figure~\ref{Rint_REF} indicates the point where the attraction from the positively biased 
$n^+$-pad on the electron layer in the inter-electrode gap 
overcomes the attractive force 
from the positive oxide charge of SiO$_2$ in the gap, which breaks the conduction channel and isolates the electrodes when the electrons are swept from the inter-electrode gap to the n$^+$-pad. At $V>V_\textrm{th,iso}$ $\rho_\textrm{int}$ becomes essentially independent 
of the $p$-stop implant, 
with absolute values 
as expected for pre-irradiated sensors \cite{Hartmann2017}.
%
%
%
\subsection{Simulated: Mixed-field and $\gamma$-irradiated sensors}
\label{MF_gammas}
\begin{figure}[htb!]
     \centering
     \subfloat[]{\includegraphics[width=.49\textwidth]{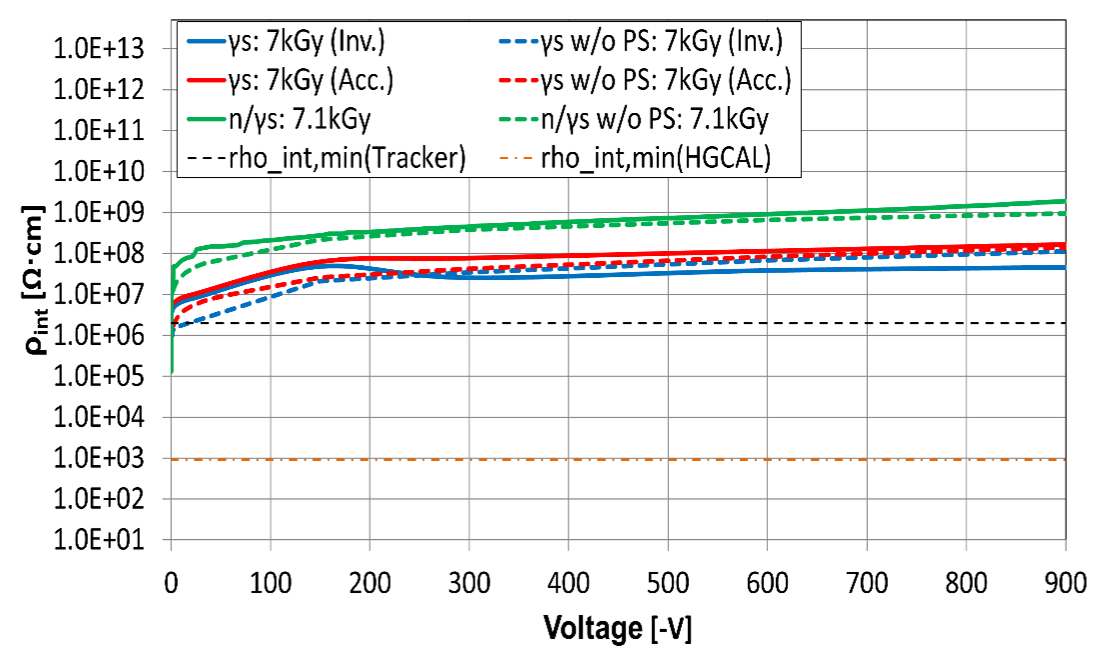}\label{RintV_7kGy}}\hspace{1mm}%
     \subfloat[]{\includegraphics[width=.495\textwidth]{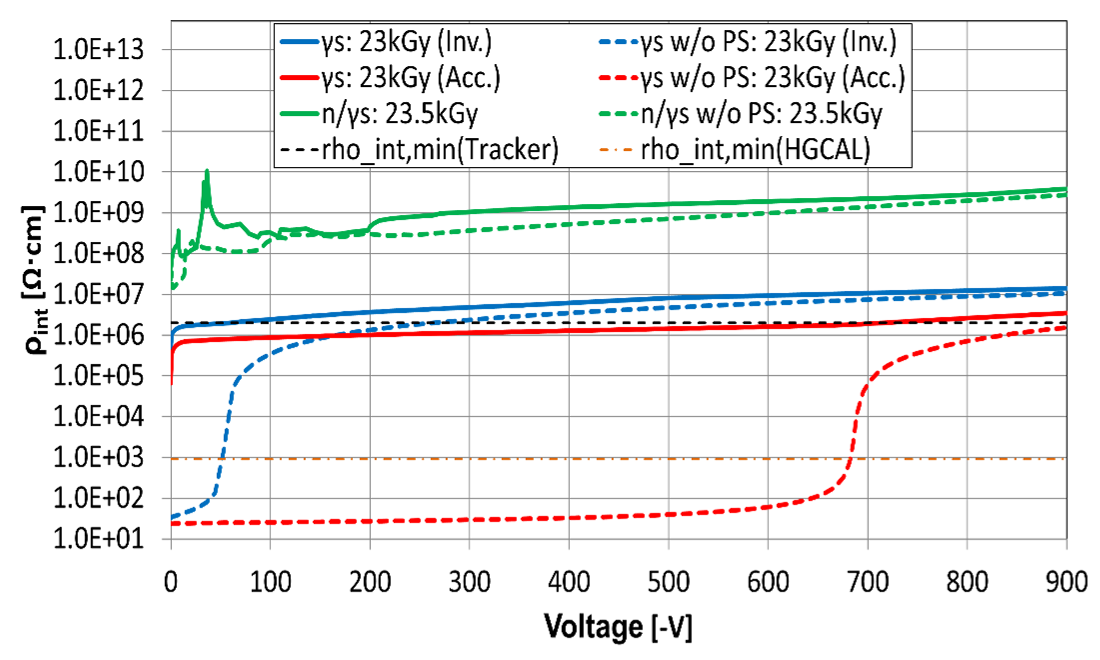}\label{RintV_23kGy}}\\
     \subfloat[]{\includegraphics[width=.495\textwidth]{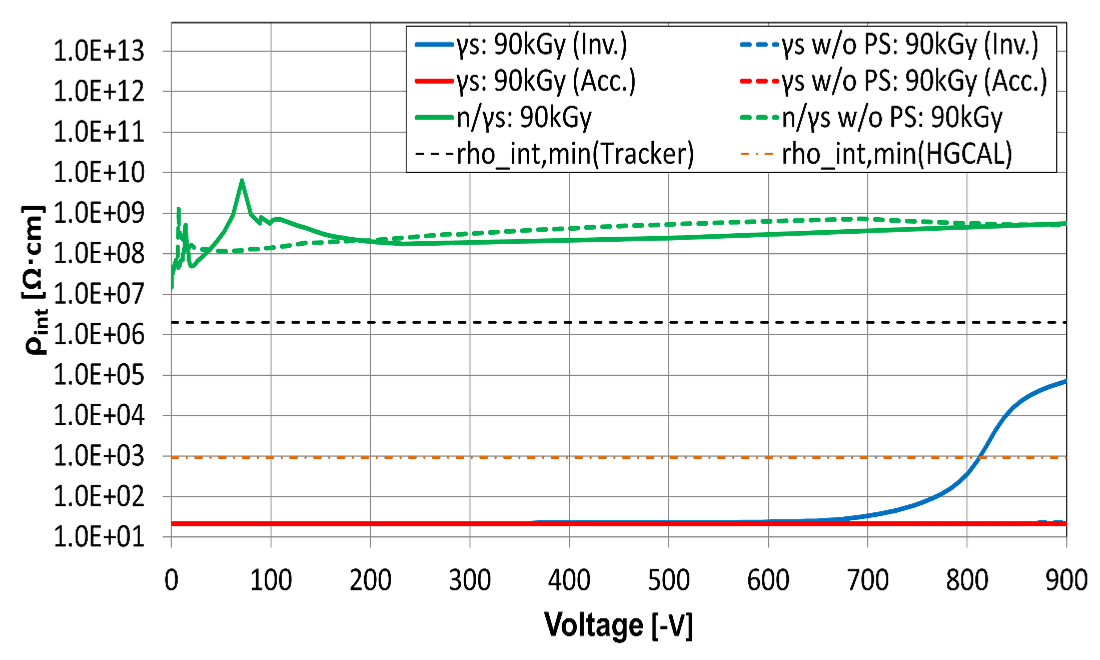}\label{RintV_90kGy}}\hspace{1mm}%
     \subfloat[]{\includegraphics[width=.495\textwidth]{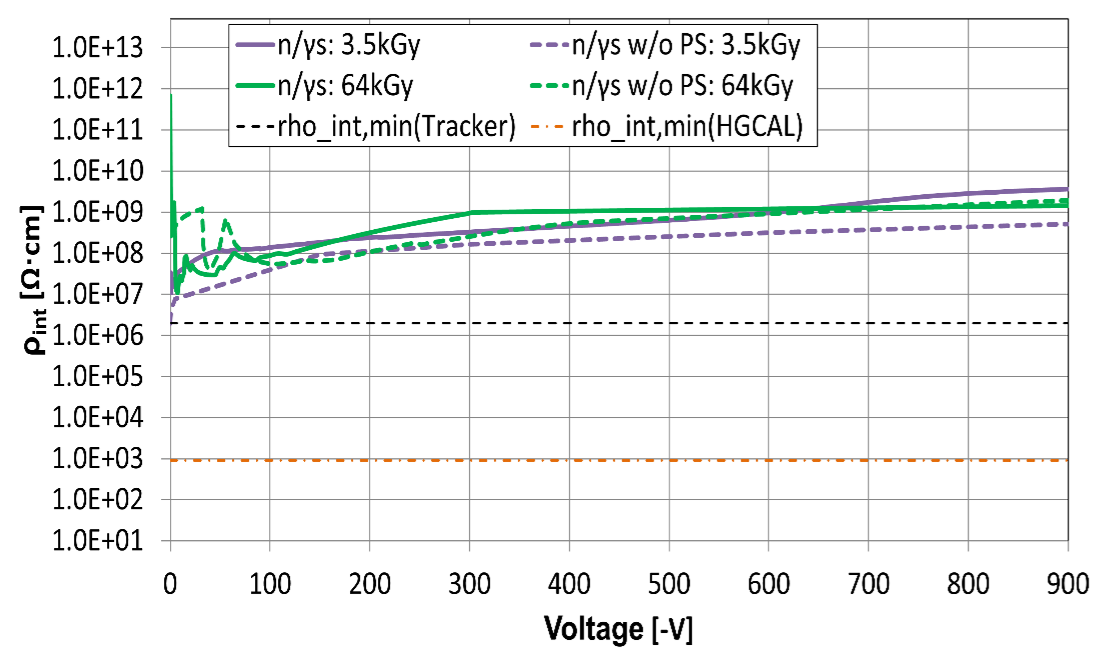}\label{RintV_35_64kGy}}
    \caption{\small Simulated evolution of $\rho_\textrm{int}$ with reverse bias voltage at $T=253~\textrm{K}$ for 
    $n$-on-$p$ sensors with and without $p$-stop implants $\gamma$- or mixed-field (n/$\gamma$) irradiated to (a) 
    $D=7.0\pm0.4~\textrm{kGy}$ and $\textrm{TID}=7.1\pm0.6~\textrm{kGy}$, (b) $D=23.0\pm1.2~\textrm{kGy}$ and $\textrm{TID}=23.5\pm1.9~\textrm{kGy}$ and (c) $D=90\pm5~\textrm{kGy}$ and $\textrm{TID}=90\pm11~\textrm{kGy}$, respectively. (d) Corresponding simulation results for the sensors mixed-field irradiated to $\textrm{TID}=3.5\pm0.3~\textrm{kGy}$ and $64\pm7~\textrm{kGy}$, respectively. Surface-state and oxide charge parameters modeling the radiation-induced surface damage are given in Tables~\ref{tabNit} and~\ref{table_Simulated}. Results obtained with parameters extracted from $CV$-sweeps started either from inversion (`Inv.') or accumulation (`Acc.') regions of the $\gamma$-irradiated MOS-capacitors are indicated in Figures~\ref{RintV_7kGy},~\ref{RintV_23kGy} and~\ref{RintV_90kGy}.
}
\label{RintV}
\end{figure}

The parameters for $N_\textrm{f}$ and $N_\textrm{it}$ from Tables~\ref{tabNit} and~\ref{table_Simulated} were used as an input for the $\rho_\textrm{int}$-simulations in Figure~\ref{RintV} for five mixed-field and three $\gamma$-irradiation TIDs 
at $T=253~\textrm{K}$, that was commonly used measurement temperature 
in previously reported $R_\textrm{int}$-studies of irradiated strip-sensors \cite{Gosewich2021_J,Gosewich2021}.

Since $\rho_\textrm{int}$ results are extracted from DC-currents in Eq.~\ref{eq2}, $f,T$-dependence of AC-measured $CV$-charac-\\
teristics of irradiated MOS-capacitors discussed in Section~\ref{TdS} is reduced to $T$-dependence. The decrease of capture and emission rates of $N_\textrm{it}$ 
with temperature 
results in lower 
surface current levels 
and thus higher absolute values of $\rho_\textrm{int}$ at voltages where pads or strips are isolated. 

As described in \cite{Peltola2023}, for the mixed-field irradiated 
sensors the 
change of the effective bulk doping ($N_\textrm{eff}$) with neutron fluence induced displacement damage was approximated by 
tuning $N_\textrm{eff}$ to obtain similar minimum-capacitance to oxide-capacitance ($C_\textrm{min}/C_\textrm{ox}$) ratios with the MOS-capacitor measurements. 

\begin{figure}[htb!]
     \centering
    \includegraphics[width=.7\textwidth]{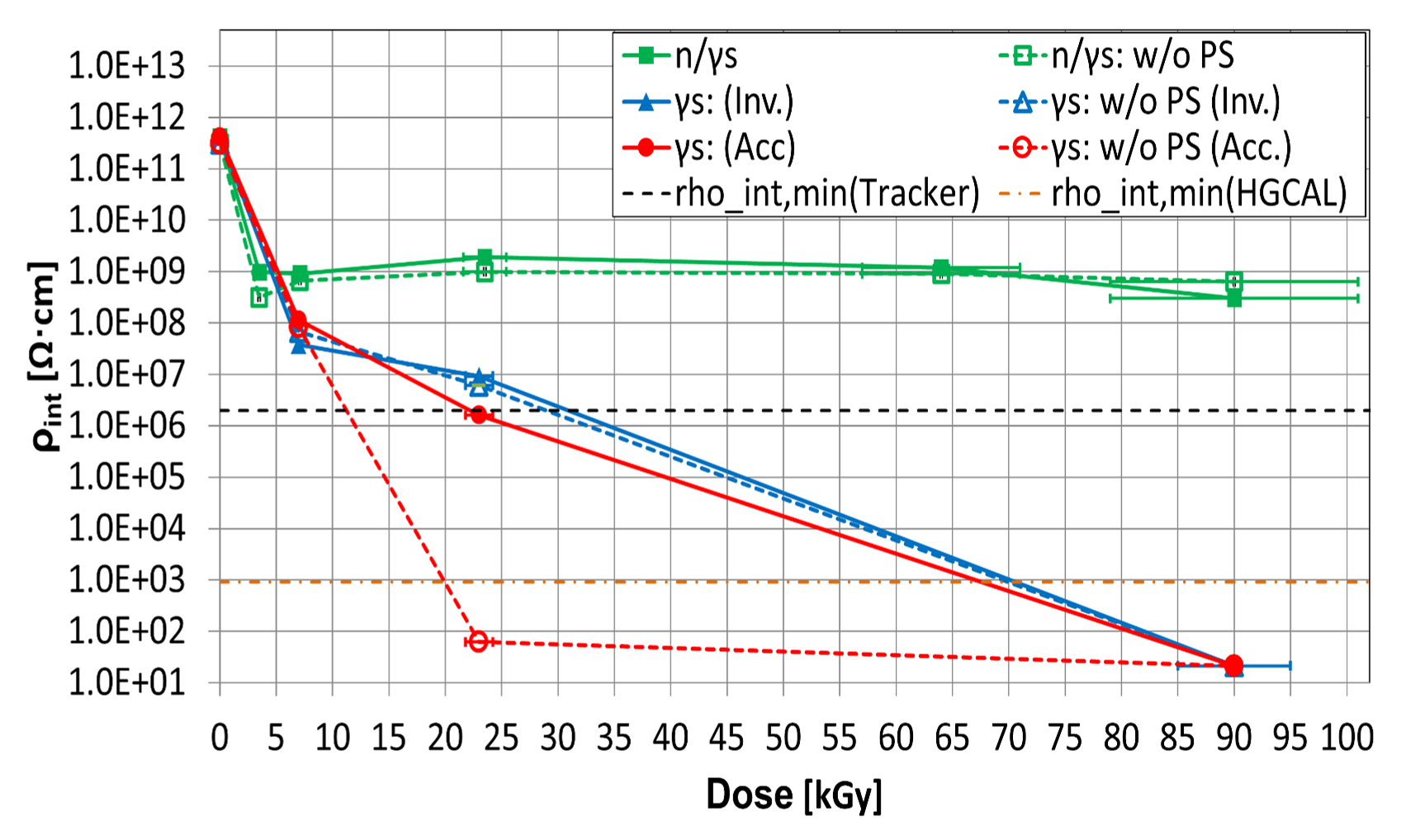}
    \caption{\small Simulated evolution of $\rho_\textrm{int}$ with dose at $T=253~\textrm{K}$ and $V_\textrm{bias}=600~\textrm{V}$, extracted from Figures~\ref{Rint_REF} and~\ref{RintV}.
}
\label{Rint_600V}
\end{figure}

Presented in Figures~\ref{RintV_7kGy}--\ref{RintV_35_64kGy}, $\rho_\textrm{int}$ for the mixed-field environment (green and violet curves) remains well above both $\rho_\textrm{int,min}$ levels at all voltages for the full TID-range from $3.5\pm0.3~\textrm{kGy}$ to $90\pm11~\textrm{kGy}$. Also, it is evident that the $p$-stop implant is essentially irrelevant for the isolation of the $n^+$-pads, which shows the beneficial impact of high introduction rate of $N_\textrm{it}$ in Figure~\ref{surfStatDens} on pad isolation. 
Corresponding results for $\gamma$-irradiated sensors in Figures~\ref{RintV_7kGy}--\ref{RintV_90kGy} (red and blue curves) display considerably lower $\rho_\textrm{int}$-levels above $7~\textrm{kGy}$ than for the mixed-field irradiated sensors with similar doses. The $p$-stop implant is now providing significant improvement on pad isolation at $D=23.0\pm1.2~\textrm{kGy}$ in Figure~\ref{RintV_23kGy} (for $CV$-sweep starting from accumulation (Acc.), red curves), while at $D=90\pm5~\textrm{kGy}$ in Figure~\ref{RintV_90kGy} limited benefit from $p$-stop starts taking place at $V>800~\textrm{V}$, where $\rho_\textrm{int}$ exceeds $\rho_\textrm{int,min}$(HGCAL) (for $CV$-sweep starting from inversion (Inv.), blue solid curve). The lower introduction rate of $N_\textrm{it}$ with $\gamma$-irradiation compared to the mixed-field in Figure~\ref{surfStatDens} correlates with the 
degrading pad-isolation performance with dose. 
Figure~\ref{Rint_600V} presents the 
evolution of $\rho_\textrm{int}$ in Figures~\ref{Rint_REF} and~\ref{RintV} with dose at 600 V (nominal operating voltage of HGCAL sensors for most of the expected fluence range \cite{Phase2}). For mixed-field irradiated sensors $\rho_\textrm{int}$ remains essentially constant after the drop of about two orders of magnitude (at $3.5\pm0.3$ kGy) from the pre-irradiated level, and at least same amount above $\rho_\textrm{int,min}$(Tracker) for the full dose range, while again $p$-stop has a negligible role on the inter-pad isolation. At $D>23~\textrm{kGy}$ $\rho_\textrm{int}$ for $\gamma$-irradiated sensors drops below $\rho_\textrm{int,min}$(Tracker) with significant benefit from $p$-stop on the pad isolation at 23 kGy (for $CV$-sweep starting from accumulation, red curves). 
At $90\pm5~\textrm{kGy}$ the pads remain shorted with and without $p$-stop (red and blue curves). The distinctly different $\rho_\textrm{int}$ performance at 23 kGy 
between parameter sets extracted from $CV$-sweeps starting either from inversion or accumulation for $\gamma$-irradiated sensors without $p$-stops will be further 
discussed in Section~\ref{StripSensors}.
Thus, the correlation of the substantially different inter-electrode isolation performances between the two irradiation types to the different introduction rates of $N_\textrm{it}$ is evident in Figures~\ref{RintV}--\ref{Rint_600V} and~\ref{surfStatDens}. 
Additionally, the general trends in the different $\rho_\textrm{int}$ performances between $\gamma$- and n/$\gamma$-irradiated sensors 
are in line with the measured results between X-ray and X-ray + n/$\gamma$-irradiated sensors with $p$-stops in \cite{Mariani2020}, where $R_\textrm{int}$ of X-ray irradiated sensors keeps decreasing with dose, while remaining essentially constant for the mixed-field irradiated sensors. 
%
%

\begin{figure}[htb!]
     \centering
     \subfloat[]{\includegraphics[width=.495\textwidth]{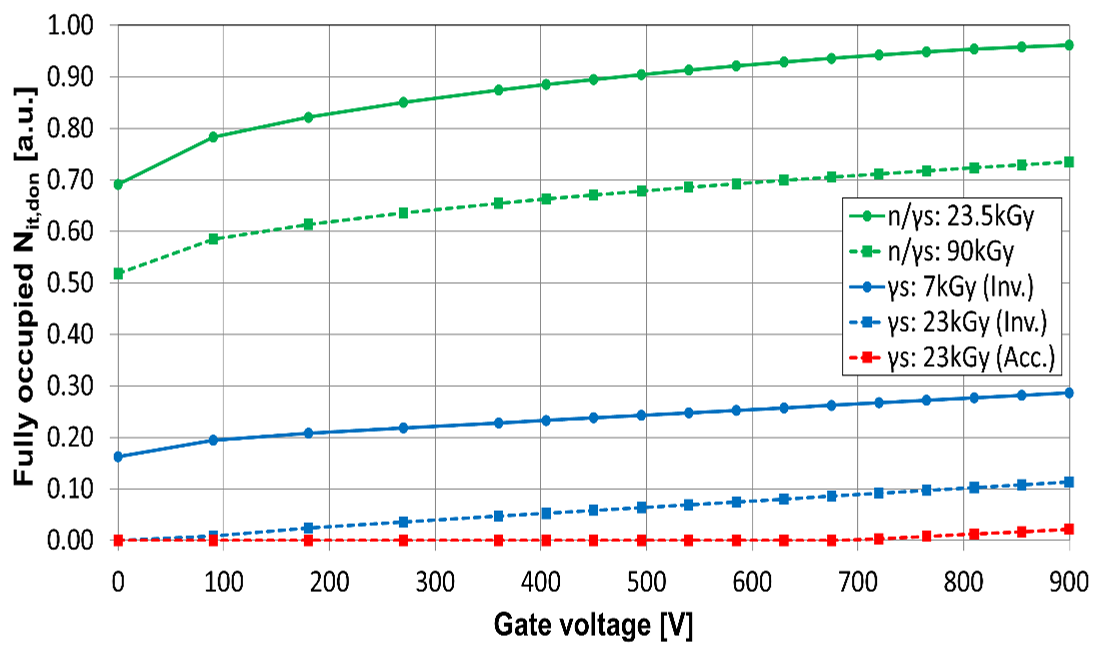}\label{fullNit_don}}\hspace{1mm}%
     \subfloat[]{\includegraphics[width=.495\textwidth]{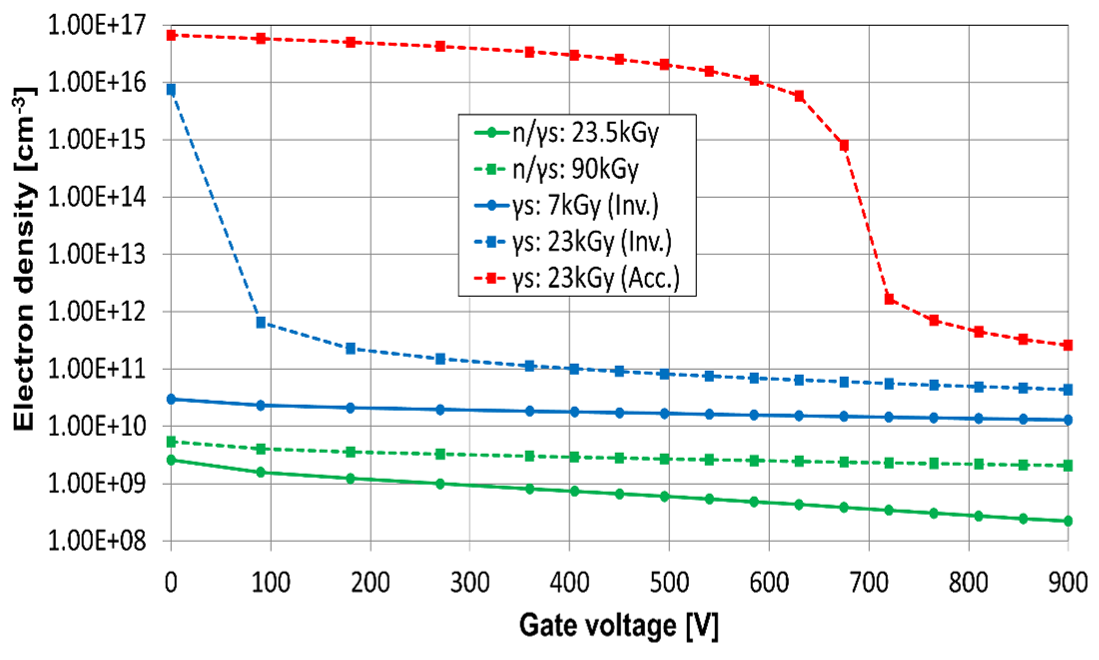}\label{eDensity}}\\
     \subfloat[]{\includegraphics[width=.495\textwidth]{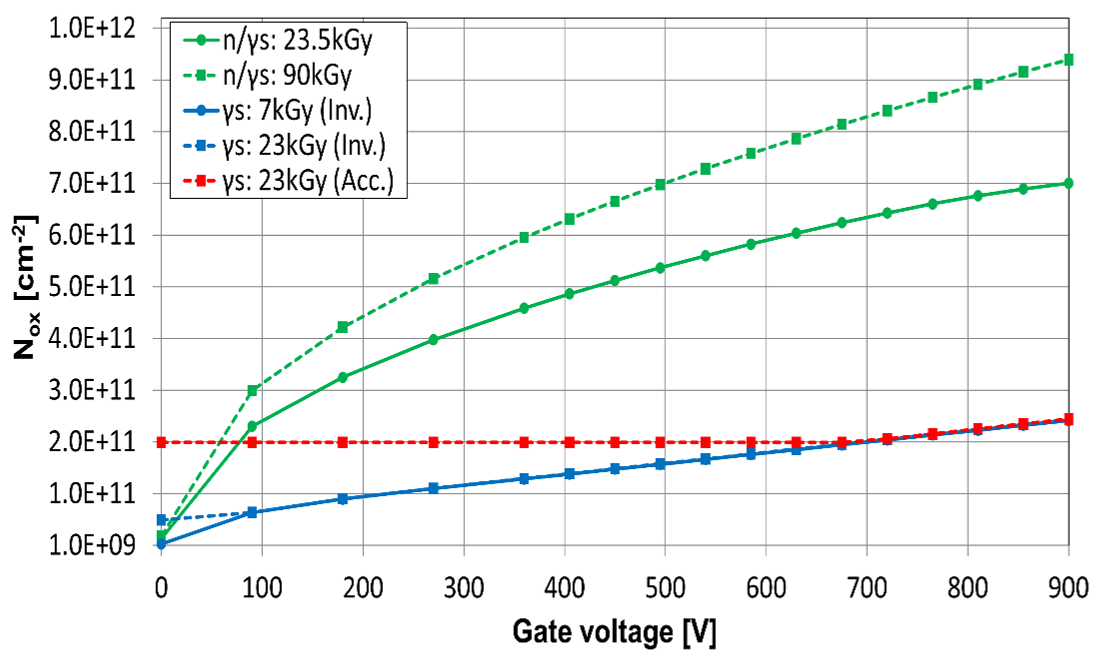}\label{NoxV}}%
    \caption{\small Simulated gate (= $n^+$-electrode) voltage dependence of 
    interface properties for high and low $\rho_\textrm{int}$ cases, with cuts at Si/SiO$_2$-interface made at mid-gap between $n^+$-implants in Figure~\ref{RintDevice1} without $p$-stop. (a) Corresponding fractions of 
    occupied $N_\textrm{it,don}$, 
(b) electron densities at the interface and (c) net $N_\textrm{ox}$ to $\rho_\textrm{int}$-curves in Figures~\ref{RintV_7kGy} ($7.0\pm0.4~\textrm{kGy}$),~\ref{RintV_23kGy} ($23.0\pm1.2$ and $23.5\pm1.9~\textrm{kGy}$) and~\ref{RintV_90kGy} ($90\pm11~\textrm{kGy}$). 
%
}
\label{Rint_mechanix}
\end{figure}

Presented in Figure~\ref{Rint_mechanix} is the comparison of the evolution of the 
Si/SiO$_2$-interface properties in the inter-electrode gap with the positive potential at the 
$n^+$-pad between selected high and low $\rho_\textrm{int}$-performance sensors (two mixed-field and two $\gamma$-irradiated sensors) in Figure~\ref{RintV}. 
Having the highest potential difference with respect to the $n^+$-electrodes, the mid-gap in Figure~\ref{Rint_mechanix} is the most revealing region on the interface, since it will be the first location where the conduction channel gets broken when the shorted electrodes 
become isolated by the bias voltage. 
Figure~\ref{fullNit_don} shows the fraction of 
occupied $N_\textrm{it,don}$ (factor $a$ in Eq.~\ref{eq1}) from their total density in Table~\ref{table_Simulated}, that reflect closely the voltage dependence of the respective $\rho_\textrm{int}$ in Figures~\ref{RintV_7kGy},~\ref{RintV_23kGy} and~\ref{RintV_90kGy}. 
For the $\gamma$-irradiated sensors, the correlation is prominent for the sensors without $p$-stop. 
This can be observed when the abrupt change from low to high values of $\rho_\textrm{int}$ at about 100 and 700 V for sensors with $23.0\pm1.2~\textrm{kGy}$ dose in Figure~\ref{RintV_23kGy} (for $N_\textrm{f}$ and $N_\textrm{it}$-parameters extracted for $CV$-sweeps from inversion or accumulation regions, respectively) is reflected as a change of factor $a$ to non-zero values in Figure~\ref{fullNit_don} (blue and red dashed curves). 
Similar but reversed-order correlation is apparent in the electron densities at the interface in Figure~\ref{eDensity}, 
with the electron density in the $23.0\pm1.2~\textrm{kGy}$ $\gamma$-dose sensors dropping by about five orders of magnitude when high $\rho_\textrm{int}$-levels are reached in Figure~\ref{RintV_23kGy}. 

Correlation between $N_\textrm{ox}$-levels in Figure~\ref{NoxV} and $\rho_\textrm{int}$ can be observed in the two green curves for mixed-field irradiated sensors, where higher positive net $N_\textrm{ox}$-values 
result in lower $\rho_\textrm{int}$ due to the increased attraction of electrons to the interface in the inter-electrode gap. While 
less pronounced, the same correlation applies also to 
the $\gamma$-irradiated sensors in Figure~\ref{NoxV} (the values of $N_\textrm{ox}$ at 900 V from the lowest to the highest $\rho_\textrm{int}$ sensor are 2.46, 2.44 and $2.43\times10^{11}~\textrm{cm}^{-2}$, respectively). 
When the pads of the $\gamma$-irradiated sensor without $p$-stop are shorted at $V<700~\textrm{V}$ in Figure~\ref{RintV_23kGy} the Si/SiO$_2$-interface in the inter-pad gap is 
filled by electrons, while hole access to the interface is restricted, leading to $a\rightarrow0$ and $b\rightarrow1$ in Eq.~\ref{eq1}. As shown by the red curve 
for 23 kGy $\gamma$-irradiation in Figure~\ref{NoxV}, net $N_\textrm{ox}$ reaches then its minimum given by $N_\textrm{ox}=N_\textrm{f}-N_\textrm{it,acc}$, which remains constant until positive potential at the $n^+$-pad becomes high enough to start removing electrons from the inter-pad gap. Hence, dynamic trends ($a>0$ and $b<1$) in the net $N_\textrm{ox}$-curves of the $\gamma$-irradiated sensors in Figure~\ref{NoxV} indicate pad isolation. However, it is evident from Figure~\ref{Rint_mechanix} that the different levels of $\rho_\textrm{int}$ are reflected only by the evolution of 
occupied $N_\textrm{it,don}$ and electron density in Figures~\ref{fullNit_don} and~\ref{eDensity}, respectively.
%
%
\subsection{Measured and simulated: X-ray and $\gamma$-irradiated sensors}
\label{StripSensors}
\begin{figure}[htb!]
     \centering
	\includegraphics[width=.7\textwidth]{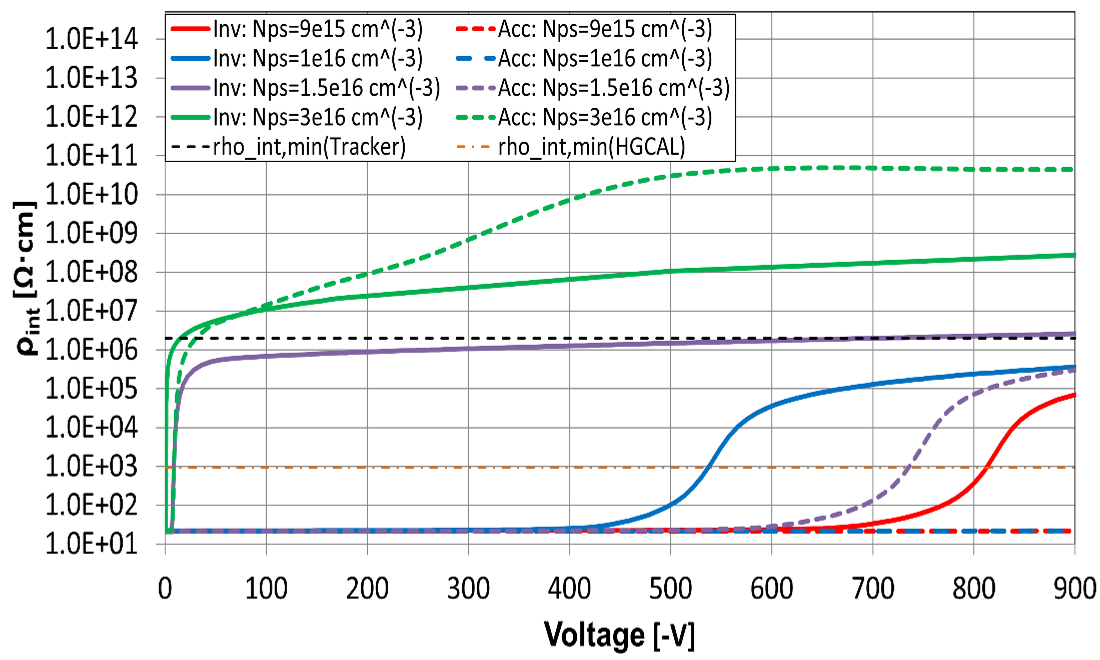}
    \caption{\small Simulated influence of $N_\textrm{ps}$ on $\rho_\textrm{int}$ at $T=253~\textrm{K}$ in a sensor $\gamma$-irradiated to $D=90\pm5~\textrm{kGy}$. Results obtained with parameters extracted from $CV$-sweeps started either from inversion (`Inv.') or accumulation (`Acc.') regions of the $\gamma$-irradiated MOS-capacitors are indicated.
}
\label{Rint_Nps}
\end{figure}

As illustrated in Figures~\ref{RintV_23kGy},~\ref{RintV_90kGy} and~\ref{Rint_600V}, $\rho_\textrm{int}$ performances of $\gamma$-irradiated sensors show considerable sensitivity to the presence of $p$-stop implant with $N_\textrm{ps}=9.0\times10^{15}~\textrm{cm}^{-3}$. Figure~\ref{Rint_Nps} shows $\rho_\textrm{int}$-results for the highest $\gamma$-irradiation dose of $D=90\pm5~\textrm{kGy}$ in Table~\ref{table_Simulated} when $N_\textrm{ps}$ is increased from its original level up to $3.0\times10^{16}~\textrm{cm}^{-3}$. It is evident that $N_\textrm{ps}$ plays a critical role in inter-electrode isolation for $\gamma$-irradiated sensors. 
In addition, $\rho_\textrm{int}$ performances display substantial differences depending on whether the parameters of $N_\textrm{f}$ and $N_\textrm{it}$ were tuned for $CV$-sweeps starting from either inversion ($\rho_\textrm{int}$(Inv.), solid curves) or accumulation ($\rho_\textrm{int}$(Acc.), dashed curves) regions. 
When $N_\textrm{ps}$ is increased to $1.0\times10^{16}~\textrm{cm}^{-3}$ $\rho_\textrm{int}\textrm{(Inv.)}>\rho_\textrm{int,min}$(HGCAL) above 500 V, and for $N_\textrm{ps}=1.5\times10^{16}~\textrm{cm}^{-3}$ $\rho_\textrm{int}\textrm{(Inv.)}$ reaches $\rho_\textrm{int,min}$(Tracker) at about 700 V. On the other hand, $\rho_\textrm{int}$(Acc.) reaches high levels for voltages below 700 V only with $N_\textrm{ps}=3.0\times10^{16}~\textrm{cm}^{-3}$, where $\rho_\textrm{int}$ for both parameter sets is above $\rho_\textrm{int,min}$(Tracker) for all voltages.

%
\begin{figure}[htb!]
     \centering
	\includegraphics[width=.7\textwidth]{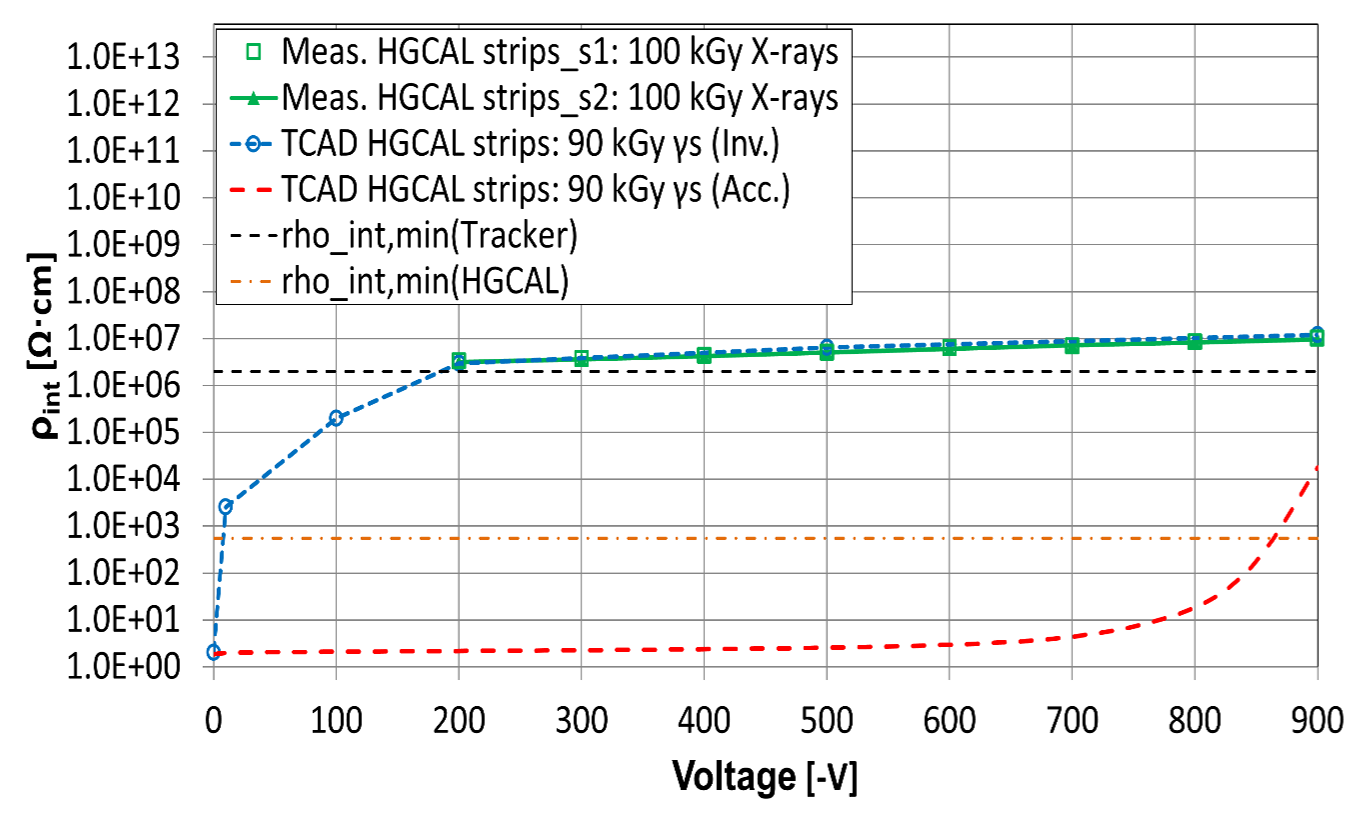}
    \caption{\small Comparison of measured and simulated $\rho_\textrm{int}$ of irradiated HGCAL test-strips at $T=253~\textrm{K}$. The two measured test-strip samples were X-ray irradiated to $D=100\pm10~\textrm{kGy}$, while the simulation applied the surface-damage parameters for $D=90\pm5~\textrm{kGy}$ from Tables~\ref{tabNit} and~\ref{table_Simulated} and $N_\textrm{ps}=1.5\times10^{16}~\textrm{cm}^{-3}$.
}
\label{Rint_MeasSim}
\end{figure}

%
To investigate which of the two aforementioned parameter sets of $N_\textrm{f}$ and $N_\textrm{it}$ tuned for $D=90\pm5~\textrm{kGy}$ models more accurately the real conditions at the Si/SiO$_2$-interface between $n^+$-electrodes, the simulations were compared to the measured $\rho_\textrm{int}$ of HGCAL test-strips in Figure~\ref{Rint_MeasSim}. The test-strips were X-ray irradiated at CERN ObeliX facility \cite{Vaiano2023} to $D=100\pm10~\textrm{kGy}$ (within uncertainty to the simulation model) and their dimensions and doping profiles were reproduced to the simulations as shown in Figures~\ref{dopeProfiles} and~\ref{RintDevice2}. Since $\rho_\textrm{int}$ performance is now strongly dependent on $N_\textrm{ps}$, its level in the test-strips was determined to be between $1.3-1.5\times10^{16}~\textrm{cm}^{-3}$ by MOSFET measurements carried out at HEPHY\footnote{Institut f{\"{u}}r Hochenergiephysik, Nikolsdorfer G. 18, Vienna, Austria.} within the HGCAL collaboration. Displayed in Figure~\ref{Rint_MeasSim}, when the upper value of the experimental $N_\textrm{ps}$ is applied to the simulation, $\rho_\textrm{int}$(Inv.) (blue dashed curve) is in close agreement with the measurement 
throughout its voltage range, while $\rho_\textrm{int}$(Acc.) (red dash) remains in the levels where the strips are shorted for the most part of the investigated $V_\textrm{bias}$-range. 

As a consequence, it is evident that the surface-damage parameter set extracted from the measurement where the $CV$-sweep was started from negative voltages (inversion region in the case of the MOS-capacitor $\gamma$-irradiated to $90\pm5~\textrm{kGy}$ in ref. \cite{Peltola2023} and Table~\ref{table_Simulated}), i.e., from the region where factor $a\rightarrow1$ in Eq.~\ref{eq1}, models more accurately the conditions at the Si/SiO$_2$-interface between reverse-biased $n^+$-electrodes after X-ray or $\gamma$-irradiation. Also, higher introduction rates of $N_\textrm{f}$ and $N_\textrm{it,acc/don}$ extracted from the inversion-region initiated $CV$-sweep in Table~\ref{table_Simulated} are closer to the real circumstances in the inter-electrode gap. 

\section{Discussion} 
\label{Discussion}
The correlation of the substantially different 
$\rho_\textrm{int}$ performances between the 
mixed-field and $\gamma$-irradiated sensors in Section~\ref{MF_gammas} to the different introduction rates of deep $N_\textrm{it}$ in Section~\ref{TdS}, 
suggests that X-ray or $\gamma$-irradiation testing of 
sensors targeted for mixed-field or hadron-dominated radiation environment 
can lead to 
overestimations on the 
degradation of the inter-electrode isolation with radiation.
Additionally, since $\rho_\textrm{int}$ was shown to be highly sensitive to the level of $N_\textrm{ps}$ in Section~\ref{StripSensors}, $\rho_\textrm{int}$-evaluations between passivation oxide variants by X-ray or $\gamma$-irradiations (e.g. possible differences in surface-state introduction rates at Si/SiO$_2$-interface) can be compromised by variations in $N_\textrm{ps}$ between sensors, as influence from both the oxide quality and $N_\textrm{ps}$ is convoluted. 
Therefore, careful assessment of the sample sensors is required to overcome the caveats associated with comparative studies on inter-electrode isolation by X-ray or $\gamma$-irradiations. 

For mixed-field irradiated sensors, the simulated $\rho_\textrm{int}$ displays high levels that remain essentially constant ($>200~\textrm{M}\Omega\cdot\textrm{cm}$ at 600 V in Figure~\ref{Rint_600V}) up to the highest dose of about 100 kGy in the study.  
The independence of the results on the presence of the $p$-stop isolation implant show the beneficial impact of the high introduction rates of deep $N_\textrm{it}$ (due to the displacement damage caused by neutrons inside the oxide and at Si/SiO$_2$-interface \cite{Amir2009}) on inter-electrode isolation. 
Thus, the results suggest that in mixed-field irradiated $n$-on-$p$ sensors the inter-electrode isolation is predominantly provided by the radiation-induced accumulation of $N_\textrm{it}$ rather than isolation implants or $V_\textrm{bias}$ (as shown in Figure~\ref{Rint_REF} and in ref. \cite{Peltola2024}, $V_\textrm{bias}$ in pre-irradiated sensor isolates $n^+$-electrodes without $p$-stop). 

The limiting factor of the reactor mixed-field irradiations in this study is the dose range. Where at the fluence of $1\times10^{16}~\textrm{n}_\textrm{eq}\textrm{cm}^{-2}$ at HGCAL the expected ionizing doses are expected to be $1.0-1.5$ MGy \cite{Phase2}, for similar fluence at the reactor radiation environment the ionizing dose is only about 100 kGy, as discussed in Section~\ref{TdS}.
However, when the reported saturation of the accumulation of $N_\textrm{f}$ and $N_\textrm{it}$ in SiO$_2$-passivated devices at X-ray doses of about 100--200 kGy \cite{Zhang2013,Moscatelli:2017sps,Moscatelli2017} is considered also for mixed-field radiation, position resolution in $n$-on-$p$ sensors without isolation implants could be expected to be maintained also in extreme radiation environments involving hadrons.   

Since no gate voltage was applied during the MOS-capacitor irradiations in \cite{Peltola2023} that produced the surface damage parameters in Tables~\ref{tabNit} and~\ref{table_Simulated}, influence of electric field on the introduction rates of $N_\textrm{it,acc/don}$ was not investigated. 
%
The proton mechanism that generates interface traps, involves protons (hydrogen ions) created by ionizing radiation in the oxide drifting to the interface with silicon to produce dangling Si-bonds (i.e. interface traps) by breaking the
hydrogenated Si-bonds at the interface. The main effect of a positive gate bias on proton mechanism is to remove the electrons thereby decreasing the fraction that recombine with the trapped holes. Conversely, a negative gate bias drives the protons away from the interface and consequently, they do not create interface traps. Decreased electron-hole recombination and increased proton reaction with the interface 
leads to higher $N_\textrm{it}$. 
Since $n^+$-pads of HGCAL sensors will be positively biased during their operation, the electron separation from holes in the inter-pad gap by the electric field could result in 
increased 
introduction rates of $N_\textrm{it}$\footnote{Electron removal from the inter-electrode gap to $n^+$-pads by reverse bias voltage is shown in \cite{Peltola2024}}. 
The proton mechanism 
is described extensively in \cite{Hjalmarson2006}.

The inflluence of electric field distribution in the inter-pad gap of the 600-V biased HGCAL sensors on the spatial uniformity of mixed-field introduced $N_\textrm{it}$ needs also to be considered. The approach of uniform $N_\textrm{it}$ along the Si/SiO$_2$-interface applied in this study might then not accurately model the real spatial distribution when surface damage is introduced to the HGCAL sensors during their operation.
As discussed in Section~\ref{MF_gammas}, the critical region for the inter-pad isolation is in the vicinity of the mid-gap. Within $\pm$10 $\upmu\textrm{m}$ from the mid-gap, for the two highest mixed-field (expected radiation type for HGCAL) fluences in the study the differences between minimum and maximum electric field values at 600 V were found to be about 4\% ($\textrm{TID}=64\pm7~\textrm{kGy}$ in Table~\ref{table_Simulated}) and 0.4\% ($\textrm{TID}=90\pm11~\textrm{kGy}$) for sensors without $p$-stop.
Thus, 
no significant variation of electric field in the region relevant to inter-pad isolation of HGCAL sensors was observed within the scope of this study. 

Finally, the choice of device boundary at the Si/SiO$_2$-interface applied in this study was tested by changing the boundary to the outer surface of SiO$_2$ with an interface with 100-$\upmu\textrm{m}$-thick layer of gas\footnote{$N_\textrm{f}=0$ and $10^{11}~\textrm{cm}^{-2}$ at the SiO$_2$/gas-interface.}, similar to the approach in \cite{Poehl2013_2}. While not identical, the inter-pad performance was observed to remain essentially the same throughout the voltage range for the new boundary, which will not affect the conclusions of this study. The comparison indicates that for a simulation study of inter-electrode isolation performance, the choice between the two boundaries is not critical.

\section{Summary and conclusions}
The parameter-tuning of surface-damage components $N_\textrm{f}$ and deep $N_\textrm{it,acc/don}$ to reproduce by TCAD simulation the $CV$-characteristics of MOS-capacitors irradiated either by $\gamma$s or a mixed field of neutrons and $\gamma$s 
showed significantly higher introduction rates of deep $N_\textrm{it,acc/don}$ for mixed-field irradiation \cite{Peltola2023}. When applied to inter-electrode resistance simulations of $n$-on-$p$ pad-sensors up to doses of about 100 kGy, higher densities of deep $N_\textrm{it,acc/don}$ 
showed correlation with higher levels of $\rho_\textrm{int}$. The 
increase of $\rho_\textrm{int}$ was found to be 
driven by the increase of the fraction of 
occupied $N_\textrm{it,don}$ and the decrease of electron density at the Si/SiO$_2$-interface in the inter-electrode gap, but not by the level of net $N_\textrm{ox}$. Additionally, comparison with measured $\rho_\textrm{int}$ of X-ray irradiated strip-sensors indicated that the 
$N_\textrm{f}$ and deep $N_\textrm{it,acc/don}$ extracted from $CV$-sweeps of $\gamma$-irradiated MOS-capacitors initiated from negative voltages (i.e., higher fraction of 
occupied $N_\textrm{it,don}$) model more accurately $\rho_\textrm{int}$ in X-ray or $\gamma$-irradiated $n$-on-$p$ sensors.

Due to the lower introduction rates of deep $N_\textrm{it,acc/don}$ relative to 
mixed field, the simulated $\rho_\textrm{int}$ of $\gamma$-irradiated sensors displayed high sensitivity to the presence and 
peak doping of a $p$-stop isolation implant above the lowest dose in the study. 
However, the presence of $p$-stop was irrelevant to the superior $\rho_\textrm{int}$ performance of the mixed-field irradiated sensors throughout the investigated dose range. The results indicate that 
while the contribution from neutrons to TID is diminutive (MCNP-simulations for the MNRC reactor indicated 5\% contribution to TID from neutrons \cite{Peltola2023}), their role in 
interface-trap introduction is decisive.

Thus, the beneficial impact of the high introduction rates of deep $N_\textrm{it,acc/don}$ on the inter-electrode isolation is evident in the results for the mixed-field radiation to the extent, where isolation implants between $n^+$-electrodes are not required to maintain uncompromised position resolution in $n$-on-$p$ sensors.
Considering both the reported pre-irradiation isolation of $n^+$-electrodes without isolation implant provided by sufficient $V_\textrm{bias}$ (less than $200~\textrm{V}$ for typical values of $N_\textrm{ox}$ in ref. \cite{Peltola2024}) and the 
apparent saturation of the accumulation of $N_\textrm{f}$ and $N_\textrm{it}$ at about $100-200$ kGy,
this configuration 
may be regarded as a possible $n$-on-$p$ sensor candidate for future HEP-experiments with radiation environments involving hadrons. However, further experimental studies are required before $n$-on-$p$ sensors without isolation implants can be considered a safe option.
Finally, similar 
number of lithography and ion-implantation steps to $p$-on-$n$ sensors would reduce the processing cost of $n$-on-$p$ sensors, while the
sensor performance of the isolation implantless configuration 
would benefit from the removal of the probability of discharges or avalanche effects due to excessive electric fields at the $p$-stops.

\label{Summary}
\FloatBarrier

\section*{Acknowledgements}
This work has been supported by the US Department of Energy, Office of Science (DE-SC0015592 and DE-SC0023690). 
We thank M. Defranchis and L. Diehl of CERN, and P. {\'A}. Dom{\'i}nguez of ETH Z{\"u}rich for providing the experimental data in the study. 
%

\appendix
\label{Appendix}
\section{MOS-capacitor modes of operation}
\label{App1}
The three modes of operation in the $CV$-characterizations of Metal-Oxide-Semiconductor (MOS) capacitors are accumulation, depletion and inversion. 
At gate voltages with $C/C_\textrm{ox}=1$, where $C$ and $C_\textrm{ox}$ are
the measured and oxide capacitances, respectively, the majority carriers (holes for $p$-bulk, electrons for $n$-bulk) are pulled to the
Si/SiO$_2$-interface, forming an accumulation layer with zero surface potential. An abrupt drop of
capacitance takes place in the depletion region, where the Si-surface is being depleted from majority
carriers and measured $C$ is now $C_\textrm{ox}$ and depletion layer capacitance in series. Thus, the measured $C$
keeps decreasing with gate voltage as the effective thickness of the depletion region, that acts as a
dielectric between the gate and the Si-substrate, increases. Depth of the depletion region reaches its
maximum, and measured $C$ its minimum, 
when most of the available minority carriers are
pulled to the Si/SiO$_2$-interface (by the positive oxide charge and by the negative gate voltage for $p$-
and $n$-bulk MOS-capacitors, respectively), 
forming an inversion layer. The depletion
region in the measured $CV$-curve is limited by the threshold voltage ($V_\textrm{th}$), where the surface potential
equals twice the bulk potential, and the flat band voltage ($V_\textrm{fb}$), where the Si energy band becomes
flat and the surface potential goes to zero \cite{Nicollian1982,Peltola2023}.

\bibliography{mybibfile}


\end{document}